\documentclass{PoS}

\title{Lattice gradient flow with tree-level $\mathcal{O}(a^4)$ improvement in pure Yang-Mills theory}

\ShortTitle{Lattice gradient flow with tree-level $\mathcal{O}(a^4)$ improvement in pure Yang-Mills theory}

\author{\speaker{Norihiko Kamata}\\
        Department of Physics, Tohoku University,\\
        E-mail: \email{kamata@nucl.phys.tohoku.ac.jp}}
        
\author{Shoichi Sasaki\\
        Department of Physics, Tohoku University,\\
        E-mail: \email{ssasaki@nucl.phys.tohoku.ac.jp}}

\abstract{
Following a recent paper by Fodor {\it et al.} (arXiv:1406.0827), we reexamine
several types of tree-level improvements on the flow action with various
gauge actions in order to reduce the lattice discretization errors in the Yang-Mills gradient flow method.
We propose two types of tree-level, $\mathcal{O}(a^4)$ improved lattice gradient flow 
including the rectangle term in both the flow and gauge action within the minimal way.
We then perform numerical simulations with the simple plaquette gauge action
for testing our proposal. Our numerical results of the expectation value of the 
action density, $\langle E(t)\rangle$, 
show that two $\mathcal{O}(a^4)$ improved flows 
significantly eliminate the discretization corrections in the small flow time $t$ regime. 
On the other hand, the values of $t^2\langle E(t)\rangle$ in the large $t$ regime, 
where the lattice spacing dependence of the tree-level term dies out as inverse powers of $t/a^2$, 
are different between the results given by two optimal flows leading 
to the same $\mathcal{O}(a^4)$ improvement
at tree level. This may suggest that non-negligible $\mathcal{O}(g^2 a^2)$ effect sets 
in the large $t$ regime, where the running coupling $g(1/\sqrt{8t})$ becomes large. 
}

\FullConference{The 33rd International Symposium on Lattice Field Theory\\
		14 -18 July 2015\\
		Kobe International Conference Center, Kobe, Japan*}
		
\usepackage{ulem}
\begin{document}
\section{Introduction}
Recently, the Yang-Mills gradient flow method \cite{Luscher:2010iy} has continued to 
develop remarkably. Indeed, this method is extremely useful for setting 
a reference scale \cite{{Luscher:2010iy},{Borsanyi:2012zs}},
definition of energy momentum tensor (EMT) 
on the lattice~\cite{{Suzuki:2013gza},{DelDebbio:2013zaa}},
calculation of thermodynamics quantities \cite{Asakawa:2013laa} and so on~\cite{Luscher:2013vga}. 
The most successful application is demonstrated in an
accurate determination of a reference scale by measuring the expectation value of the 
action density $E(t, x)$. However, there is still room for improvement with respect to the lattice 
gradient flow, where some lattice artifacts are found to 
be non-negligible~\cite{{Fodor:2014cpa},{Ramos:2014kka}}. Therefore,
it is important to understand how to practically reduce the effects of the lattice artifacts,
especially lattice discretization errors as a result of the finite lattice spacing $a$.

At tree level in the gauge coupling, 
lattice discretization effects on the expectation value $\langle E(t) \rangle$ 
had been already studied in the last year~\cite{{Fodor:2014cpa},{Ramos:2014kka},{Ramos:2015baa}}. 
According to their results, tree-level discretization errors become large in the small flow-time ($t$) 
regime as inverse powers of $t/a^2$. 
This tendency is problematic when we construct the lattice EMT operator
using the Yang-Mills gradient flow and then calculate thermodynamics quantities such as trace anomaly 
and entropy density following Suzuki's proposal~\cite{Suzuki:2013gza}.
The idea of Suzuki method is based on the fact that flowed observables, which live on $4 + 1$ 
dimensional space, can be expanded by a series of the expectation values of the ordinary four-dimensional operator 
in powers of the flow time $t$, (so-called ``small $t$ expansion'') \cite{Luscher:2013vga}. 
Therefore, it is important to control tree-level lattice discretization errors 
on the action density $E(t, x)$, which is a key ingredient to evaluate the trace anomaly term in the lowest-order 
formula of the new EMT construction~\cite{Suzuki:2013gza}.

In this context, we discuss what is an optimal combination of choices of the flow, the gauge action
and the action density in line with the tree-level improvement for the lattice gradient flow~\cite{Fodor:2014cpa}. 
A simple idea of achieving $\mathcal{O}(a^2)$ improvement 
was considered by FlowQCD Collaboration~\cite{private}. 
The appropriate weighted average of the values $t^2\langle E(t)\rangle$, which are obtained 
by the plaquette and clover lattice versions of $E(t,x)$, can easily cancel their $\mathcal{O}(a^2)$ corrections.
The weight combination was determined at tree-level in Refs.~\cite{Fodor:2014cpa} and \cite{Ramos:2014kka}. 
We thus develop this idea to achieve tree-level $\mathcal{O}(a^4)$ improvement
using the flow action with both the plaquette and rectangle in this paper. 

\section{The Yang-Mills gradient flow and its tree-level discretization effects}
Let us briefly review the Yang-Mills gradient flow and its tree-level discretization corrections. 
The Yang-Mills gradient flow is a kind of diffusion equation where the gauge fields $A _\mu(t, x)$ 
evolves smoothly as a function of fictitious time $t$. It is expressed by the following equation,
\begin{equation}
\frac{dA_{\mu} (t, x)}{dt} = -\frac{\delta S_{YM}[A]}{\delta A_{\mu}(t, x)},
\end{equation}
where $S_{YM}[A]$ denotes the pure Yang-Mills action defined in terms of the flowed gauge 
fields $A_{\mu}(t, x)$. 
The initial condition of the flow equation at $t=0$, $A_\mu(0, x)$, 
are supposed to correspond to the gauge fields of the 4-dimensional pure Yang-Mills theory. 
Through above flow equation, the gauge fields can be smeared out over the sphere with a
radius roughly equal to $\sqrt{8t}$ in the ordinary 4-dimensional space-time.
One of the major benefits of the Yang-Mills gradient flow, is that correlation functions of the flowed gauge fields 
$A _\mu(t, x)$ has no ultraviolet (UV) divergence for a positive flow time ($t>0$) under standard renormalization~\cite{{Luscher:2010iy},{Luscher:2011bx}}. 

To see this remarkable feature, let us consider a specific quantity, like the action density $E(t,x)$ that is defined by $E(t,x) = \frac{1}{2}{\rm Tr}\{G_{\mu \nu}(t,x) G_{\mu \nu}(t,x)\}$. 
Here, the field strength of the the flowed gauge fields is given by $G_{\mu \nu}=\partial_\mu A_{\nu}-\partial_\nu A_{\mu}+[A_\mu, A_\nu]$ ($\mu, \nu= 1, 2, 3, 4$) 
in the continuum expression.
Taking the smaller value of $t$ implies the consideration of high-energy behavior of the theory. 
Therefore, the vacuum expectation of $E(t,x)$ in the small $t$ regime, where the gauge coupling becomes small, 
can be evaluated in perturbation theory. The dimensionless combination $t^2\langle E (t) \rangle$
was calculated at the next-to-leading-order in powers of the renormalized coupling at a scale of $1/\sqrt{8t}$
in the $\overline{\rm MS}$ scheme~\cite{Luscher:2010iy} as below,
\begin{equation}
\label{eq:flowed_energy}
t^2 \langle E (t) \rangle = \frac{3(N_C^2 -1) g(1/\sqrt{8t})^2}{128 \pi ^2} \Bigl[ 1 + \bar{c}_1 g(1/\sqrt{8t})^2 + {\mathcal O}(g(1/\sqrt{8t})^4) \Bigr],
\end{equation}
where $\bar{c}_1$ is given as $N_C(11\gamma _E /3 +52/9 -3\ln{3})/16\pi ^2 $ 
with Euler's constant $\gamma _E$
for the number of colors $N_C$. Unlike the ordinary 4-dimensional gauge theory, 
Eq.~(\ref{eq:flowed_energy}) has no divergence term proportional to $1/\varepsilon$ at this order. 
This UV finiteness have been proved not only for the above particular quantity at this given order, 
but also for any correlation functions composed of the flowed gauge fields at 
all orders of the gauge coupling~\cite{Luscher:2011bx}. 

The lattice version of $t^2 \langle E (t) \rangle$ obtained in numerical simulations
shows a monotonically increasing behavior as a function of the flow time $t$ and 
also good scaling behavior with consistent values of the continuum perturbative calculation 
(\ref{eq:flowed_energy}) that suggests the presence of the proper continuum limit~\cite{Luscher:2010iy}. 
The observed properties of $\langle E (t)\rangle$
offer a new reference scale $t_0$, which is determined 
by a relation of $t_0^2 \langle E (t_0) \rangle = 0.3$~\cite{Luscher:2010iy}.

Tree-level discretization errors of $t^2 \langle E (t) \rangle$ had been already studied 
in Refs.~\cite{Fodor:2014cpa} and \cite{Ramos:2014kka}. 
According the paper~\cite{Fodor:2014cpa}, the lattice version of $t^2 \langle E(t) \rangle$ can be expanded 
in a perturbative series in the bare coupling $g_0$ as
\begin{equation}
\label{eq:flowed_energy_lattice}
t^2 \langle E (t) \rangle_{\rm{lat}} = \frac{3(N_C^2 -1) g_0^2}{128 \pi ^2} \Bigl[ C(a^2/t) + \mathcal{O}(g_0^2) \Bigr].
\end{equation}
The lattice spacing dependence of tree-level contribution appears in the first term, which
is classified by powers of $a^2/t$ as $C(a^2/t) = 1 +\sum _m ^{\infty} C_{2m} \cdot a^{2m} /t^m$.
The second contribution of $\mathcal{O}(g_0^2)$ represents quantum corrections beyond tree-level.
Determinations of the coefficients $C_{2m}$ depend on three building blocks: (1) a choice of the 
lattice gauge action for the configuration generation (2) a choice of the lattice version of the action 
density (3) a choice of the lattice gauge action for the flow action. In Ref.~\cite{Fodor:2014cpa}, 
the $\mathcal{O}(a^{2m})$ correction terms have been determined up to $C_8$ for various cases of
three building blocks.

For clarity, we will hereafter use a word of {\it ``X flow''}, when we adopt {\it the X gauge action for the flow}. 
For examples, we call {\it the Wilson flow} and {\it the Iwasaki flow} for choices of the Wilson and Iwasaki
gauge actions for the flow.  

\section{Tree-level $\mathcal{O}(a^4)$ improved gradient flow}
Following the tree-level improvement program proposed by Fodor {\it et al.}~\cite{Fodor:2014cpa}, 
we consider improvements of the lattice gradient flow within choices of 
two different rectangle coefficients $c_g$ for the configuration generation and $c_f$
for the flow. First of all, we describe a simple method for tree-level $\mathcal{O}(a^2)$ improvement, 
which was originally proposed by FlowQCD Collaboration~\cite{private}. 
Let us consider $C_2$ coefficient of the $\mathcal{O}(a^{2})$ correction term
with both the plaquette- and clover-type definitions of the action density $E(t, x)$.
The $C_{2}$ coefficients are given for the plaquette ($C_{2p}$) 
and clover ($C_{2c}$) as below~\cite{Fodor:2014cpa}
\begin{eqnarray}
C_{2p} = 2c_f + \frac{2}{3}c_g + \frac{1}{8}, \quad C_{2c} = 2c_f + \frac{2}{3}c_g - \frac{1}{24}.
\end{eqnarray}
Clearly, $C_{2p}\neq C_{2c}$ with the fixed $c_g$ and $c_f$. Therefore,
in order to eliminate tree-level $\mathcal{O}(a^2)$ effects, one can simply 
take a linear combination of two observables, which gives the corresponding $C_{2}$ coefficient as
$C_{2mix} = \alpha_m  C_{2p} + \beta_m C_{2c}$.
Appropriate combination factors $\alpha_m$ and $\beta_m$ can be determined under the 
condition of $C_{2mix}=0$ with the normalization $\alpha_m + \beta_m = 1$ 
so that the coefficient of the leading term is unity
and then obtained as
\begin{equation}
\label{eq:weight}
 \alpha_m = 1 - 6\biggl(2c_f + \frac{2}{3}c_g + \frac{1}{8} \biggr),\  \beta_m = 6\biggl(2c_f + \frac{2}{3}c_g + \frac{1}{8} \biggr),
\end{equation}
which can eliminate $C_{2mix}$ at any choice of $c_g$ and $c_f$.
Therefore, the following linear combination
\begin{equation}
\label{eq:combination}
\alpha_m\langle E_{\rm plaq}(t)\rangle + \beta_m\langle E_{\rm clover}(t)\rangle
\end{equation}
have no tree-level $\mathcal{O}(a^2)$ corrections.
This idea is quite simple as can be seen for the case of $c_g=c_f=0$, where a weighted average of two observables, $\frac{1}{4}\langle E_{\rm plaq}(t)\rangle + \frac{3}{4}\langle E_{\rm clover}(t)\rangle$, 
would achieve 
tree-level $\mathcal{O}(a^{2})$ improvement~\footnote{Please note that this linear combination 
is different from 
tree-level $\mathcal{O}(a^{2})$ improved ``operator'',
$\frac{4}{3} E_{\rm plaq}(t,x) - \frac{1}{3} E_{\rm clover}(t,x)$, proposed by Ramos and Sint~\cite{Ramos:2015baa}.}.

Next, we would like to develop this idea to achieve tree-level $\mathcal{O}(a^{4})$ improvement.
Taking a linear combination of $\alpha_m\langle E_{\rm plaq}(t)\rangle + \beta_m\langle E_{\rm clover}(t)\rangle$,
the corresponding $C_4$ coefficient of the $\mathcal{O}(a^{4})$ correction term are given by
\begin{equation}
\label{eq:order_4th}
 C_{4mix} = \biggl[1 - 6\biggl(x + \frac{2}{3}c_g\biggr) \biggr]C_{4p} + 6\biggl(x + \frac{2}{3}c_g\biggr) C_{4c},
\end{equation}
where $C_{4p}$ and $C_{4c}$ represent the $C_{4}$ coefficients given for the plaquette 
and clover. Here we introduce $x = 2c_f + 1/8$ for simplicity. 
According to Ref.~\cite{Fodor:2014cpa}, both $C_{4p}$ and $C_{4c}$ are
polynomial functions of $x$. When the value of $c_g$ is fixed, Eq.~(\ref{eq:order_4th}) reduces simple quadratic functions of $x$. Therefore, one can find two kinds of the optimal flow action ($c_f$) that can eliminate $C_{2mix}$ and $C_{4mix}$ coefficients simultaneously.

Throughout this paper we only consider the Wilson plaquette gauge action for the configuration generation ($c_g=0$). 
In this case, $C_{4p}$ and $C_{4c}$ are given by
\begin{eqnarray}
C_{4p} = \frac{57}{32}x^2 - \frac{25}{128}x + \frac{41}{2048}, \quad C_{4c} =  \frac{57}{32}x^2 - \frac{353}{640}x + \frac{401}{10240}.
\end{eqnarray}
In order to eliminate $C_{4mix}$, we should solve the following quadratic equation
\begin{eqnarray}
C_{4mix} = - \frac{57}{160}x^2 - \frac{103}{1280}x + \frac{41}{2048} = 0,
\end{eqnarray}
which leads to two kinds of the optimal coefficients $c_f$, 
\begin{eqnarray}
\label{eq:opt_coeff}
  c_{f1} =  -0.250261, \quad  c_{f2} =  +0.012323.
\end{eqnarray}
The first solution $c_{f1}$ is close to the rectangle coefficient of the Iwasaki gauge 
action ($c_f = -0.331$), while the second one $c_{f2}$ is very close to the zero. 
Therefore we call the former type flow as ``Iwasaki-like flow'' and the latter one as ``Wilson-like flow'' 
for convenience.  

\section{Numerical results}
We perform numerical simulations to test the tree-level $\mathcal{O}(a^4)$ improved method, 
which is proposed in the previous section. 
As summarized in Table \ref{tab:set-up}, we generate three ensembles using the same lattice set up
with the Wilson plaquette gauge action as in the original work of the Wilson flow done 
by L\"uscher~\cite{Luscher:2010iy}. 
For the lattice gradient flow, we use four types of the flow action: Wilson, Iwasaki, Wilson-like, and 
Iwasaki-like flows and evaluate $t^2 \langle E (t) \rangle$ by both the plaquette- and clover-type definitions.
To eliminate $\mathcal{O}(a^2)$ or $\mathcal{O}(a^4)$ corrections from the observable of 
$\langle E (t) \rangle$, we take the appropriate linear combinations of $\langle E_{\rm plaq}(t) \rangle$
and $\langle E_{\rm clover}(t) \rangle$ according to Eqs.(\ref{eq:weight}), (\ref{eq:combination}) 
and (\ref{eq:opt_coeff}).
For our 6 types of the lattice gradient flows, 
the $\mathcal{O}(a^2)$, $\mathcal{O}(a^4)$ and $\mathcal{O}(a^6)$ corrections
terms $C_{2,4,6}$ are summarized in Table~\ref{tab:Tree-level lattice effects}.

\begin{table}[t]
\begin{center}
\caption{
Simulation parameters of three ensembles.
The values of the Sommer scale $r_0$ and lattice spacing $a$ are
taken from Ref.~\cite{Necco:2001xg} and  Ref.~\cite{Luscher:2010iy}, respectively.
\label{tab:set-up}
}
\begin{tabular}{lccccc}
\hline
\hline
 $\beta = 6/g_0^2$ (Action) & $r_0/a$ & $a$ [fm] & ($L^3 \times T$) & $\sim La$ [fm] & Statistics \cr
\hline
5.96 (Wilson)& 5.002 & 0.0999(4)  & $24^3\times 48$ & 2.40 &  100\cr
6.17 (Wilson)& 7.061 & 0.0710(3)  & $32^3\times 64$ & 2.27 &  100\cr
6.42 (Wilson)& 10.00 & 0.0498(3)  & $48^3\times 96$ & 2.39 &  100\cr
\hline
\hline
\end{tabular}
\end{center}
\end{table}
\begin{table}[b]
\begin{center}
\caption{
Values of the coefficients in the tree-level $\mathcal{O}(a^2)$,
$\mathcal{O}(a^4)$ and $\mathcal{O}(a^6)$ terms of
$t^2\langle E\rangle$. 
\label{tab:Tree-level lattice effects}
}
\begin{tabular}{lcccc}
\hline
\hline
types of calculation & types of $\langle E\rangle$ & $C_2$ & $C_4$ & $C_6$\cr
\hline
unimproved Wilson flow & clover &  $-0.0417$      &$-0.0020 $ &$-0.0002$ \cr
unimproved Iwasaki flow& clover &  $-0.7037$      &$+0.8490$ &$-1.5093$ \cr
$\mathcal{O}(a^2)$-imp Wilson flow & plaq-plus-clover & 0 & $-0.0044$& $+0.0014$ \cr
$\mathcal{O}(a^2)$-imp Iwasaki flow & plaq-plus-clover& 0 & $-0.0395$& $+0.1362$ \cr
$\mathcal{O}(a^4)$-imp Wilson-like flow  & plaq-plus-clover& 0 &0& $+0.0004$ \cr
$\mathcal{O}(a^4)$-imp Iwasaki-like flow & plaq-plus-clover& 0 &0& $+0.0272$ \cr
\hline
\hline
\end{tabular}
\end{center}
\end{table}

\begin{figure}[t]
\includegraphics[width=.33\textwidth,clip]{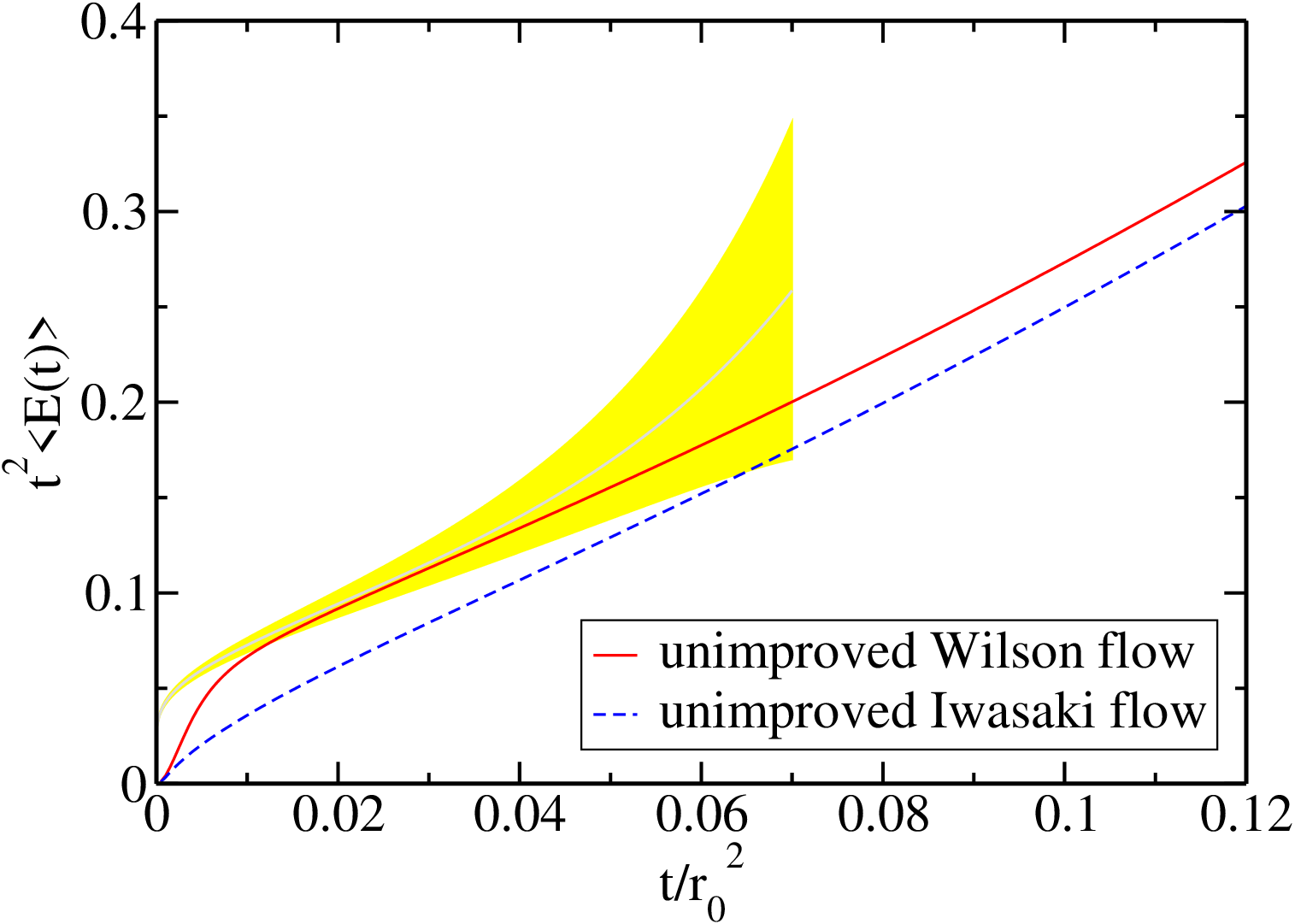}
\includegraphics[width=.33\textwidth,clip]{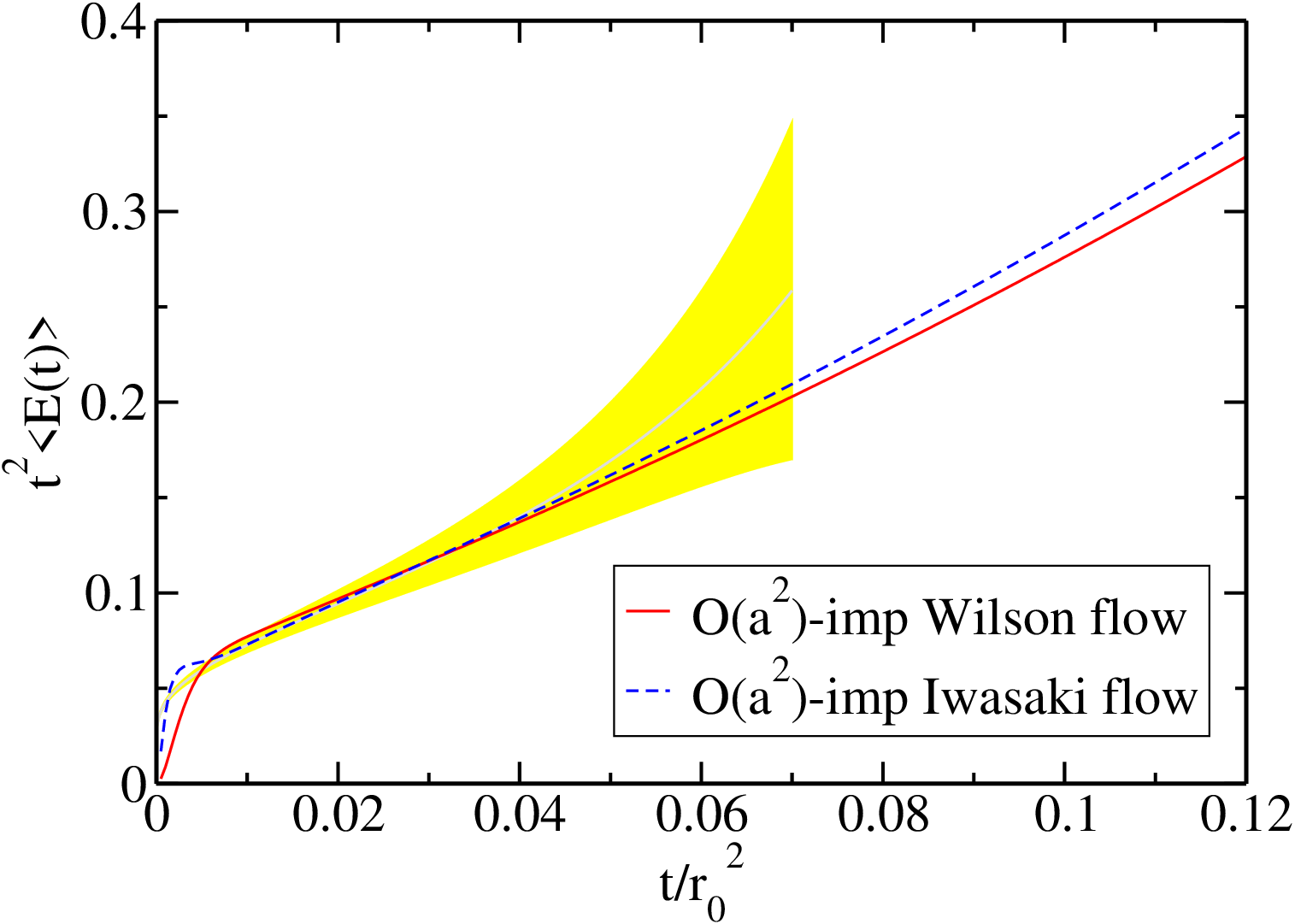}
\includegraphics[width=.33\textwidth,clip]{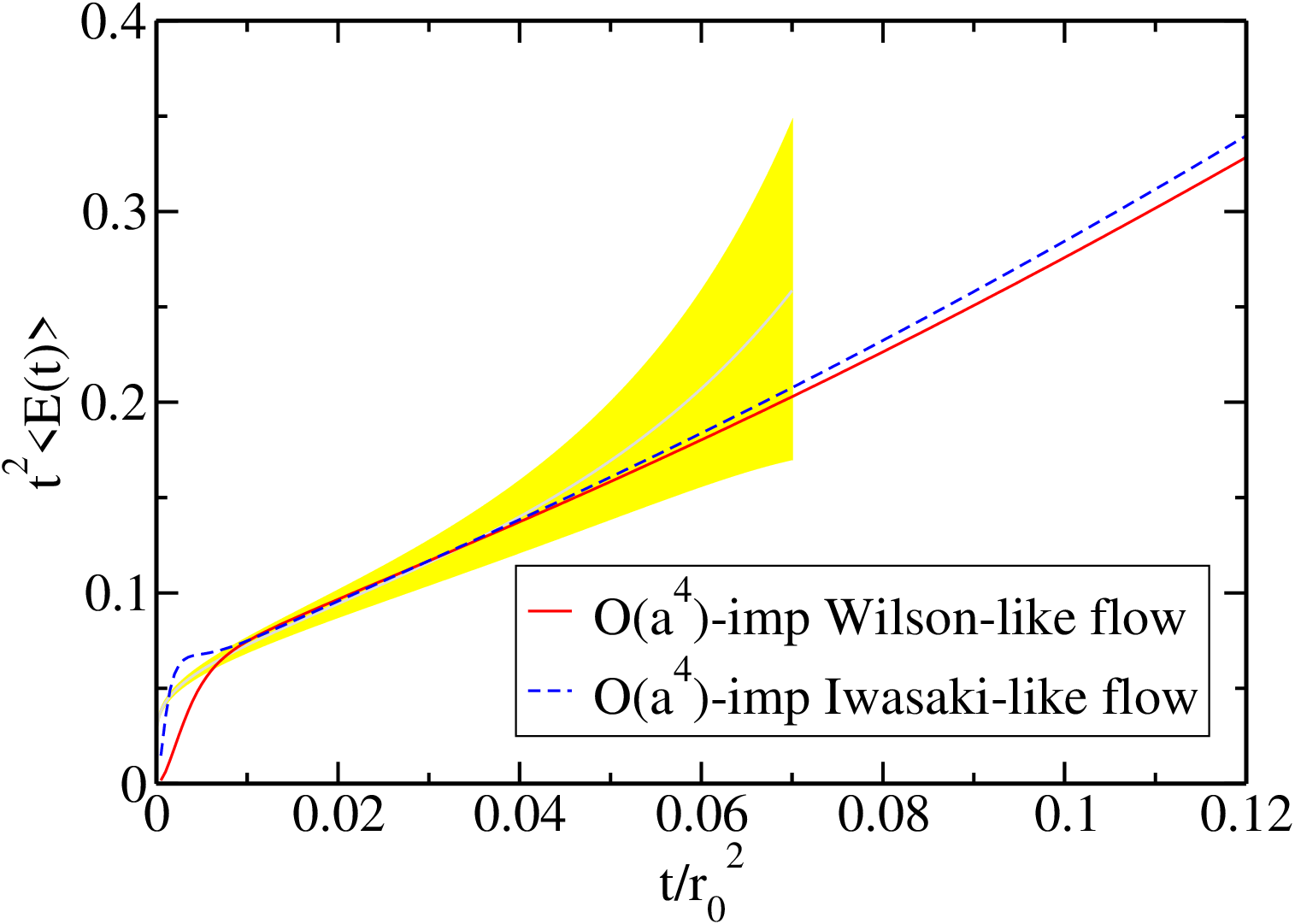}
 \caption{The behavior of $t^2 \langle E (t) \rangle$ as functions of $t/r_0^2$ at $\beta = 6.17$. 
Three panels show results of unimproved Wilson (solid) and Iwasaki (dashed) flows in the left panel, 
$\mathcal{O}(a^2)$-improved Wilson (solid) and Iwasaki (dashed) flows in the central panel 
and two types of the optimal flow, the Wilson-like (solid) and Iwasaki like (dashed) flows, 
leading to $\mathcal{O}(a^4)$ improvement in the right panel. In each panel, 
the yellow shaded bands corresponds to the continuum perturbative calculation~\cite{Luscher:2010iy}. }
 \label{fig:beta617}
 \end{figure}

In Fig.~\ref{fig:beta617}, we first show how our proposal of tree-level $\mathcal{O}(a^4)$ improvements
works well in $t$-dependence of $t^2 \langle E (t) \rangle$ 
calculated at $\beta = 6.17$.
Three panels show results of unimproved flows (left), 
$\mathcal{O}(a^2)$-improved flows (center)
and $\mathcal{O}(a^4)$-improved flows (right).
In each panel, a red solid curve is obtained from the Wilson-type flow, while a blue dashed curve
is given by the Iwasaki-type flow. The yellow shaded band represents the continuum perturbative calculation
using the next-to-leading formula (\ref{eq:flowed_energy}) with the same prescription of the running coupling 
in Ref.~\cite{Luscher:2010iy}. 

For the unimproved case (left panel), it is found that the Iwasaki flow result is away from the Wilson flow result and
the continuum perturbative calculation. However, both tree-level $\mathcal{O}(a^2)$ and $\mathcal{O}(a^4)$
improvements indeed improve results obtained from the Iwasaki-type flow significantly.  
Even for the Wilson-type flows, the improvements become visible in the smaller $t$ regime up to $t/r_0^2\approx 0.01$. Furthermore, it is observed that in the range of $0.01< t/r_0^2 < 0.05$, 
curves obtained from both the Wilson-type and Iwasaki-type flows almost coincide. This tendency is likely to be
 strong in results of the tree-level $\mathcal{O}(a^4)$ improved flows especially toward the smaller value of $t$.
This indicates that the tree-level discretization errors, which may dominant in the small $t$ regime, 
are well controlled by our proposal. However, in the large $t$ regime ($t/r_0^2>0.05$), 
the difference between results from the Wilson-type flow ($c_f\approx 0$) and the Iwasaki-type flow ($c_f\approx -0.3$) becomes evident and also increases for a larger value of $t$. It is worth mentioning that at tree-level, the higher order corrections  become negligible in the large $t$ regime due to powers of $a^2/t$.  

\begin{figure}[t]
\includegraphics[width=.33\textwidth,clip]{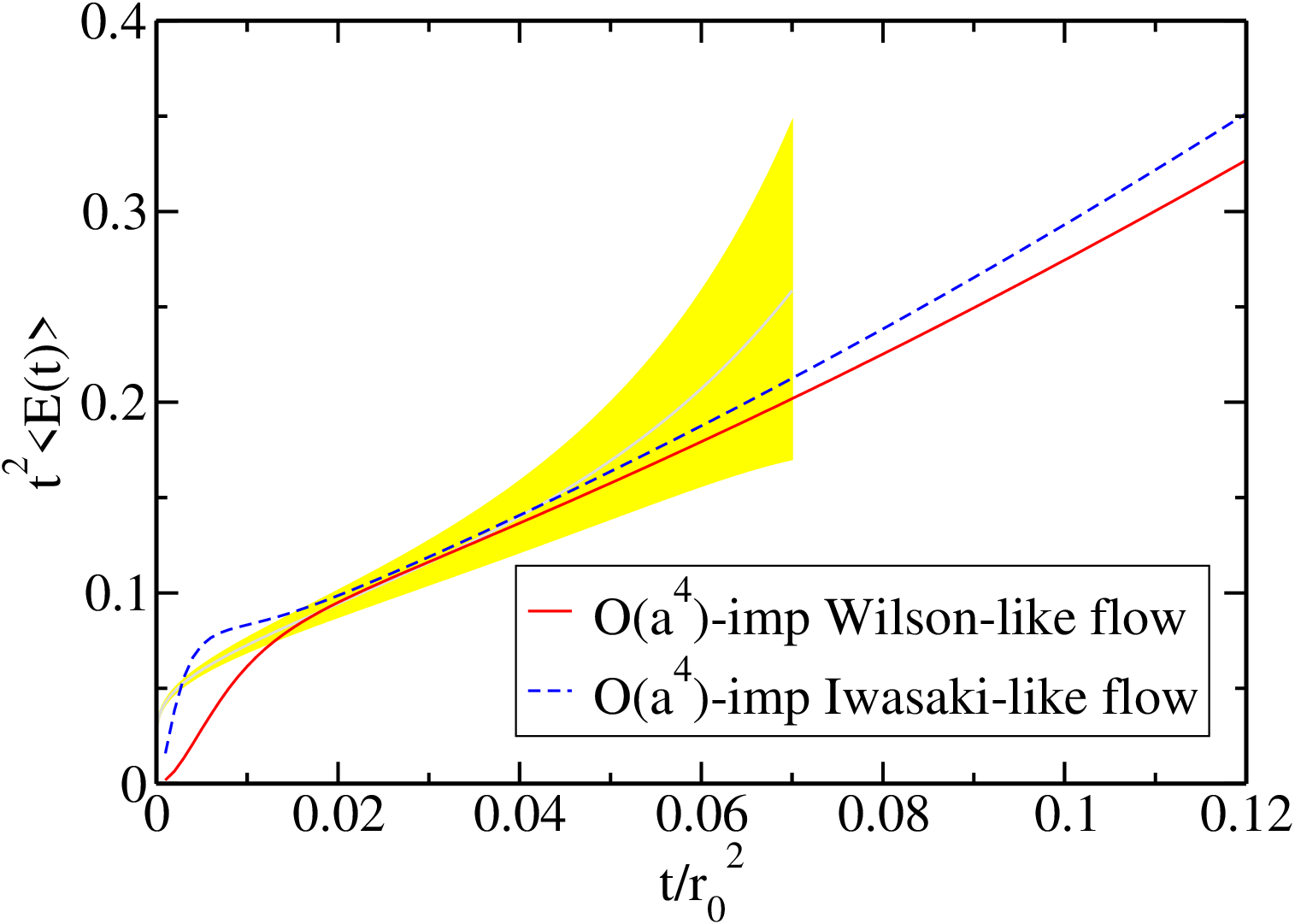}
\includegraphics[width=.33\textwidth,clip]{./eval_beta617_o4_fin.eps}
\includegraphics[width=.33\textwidth,clip]{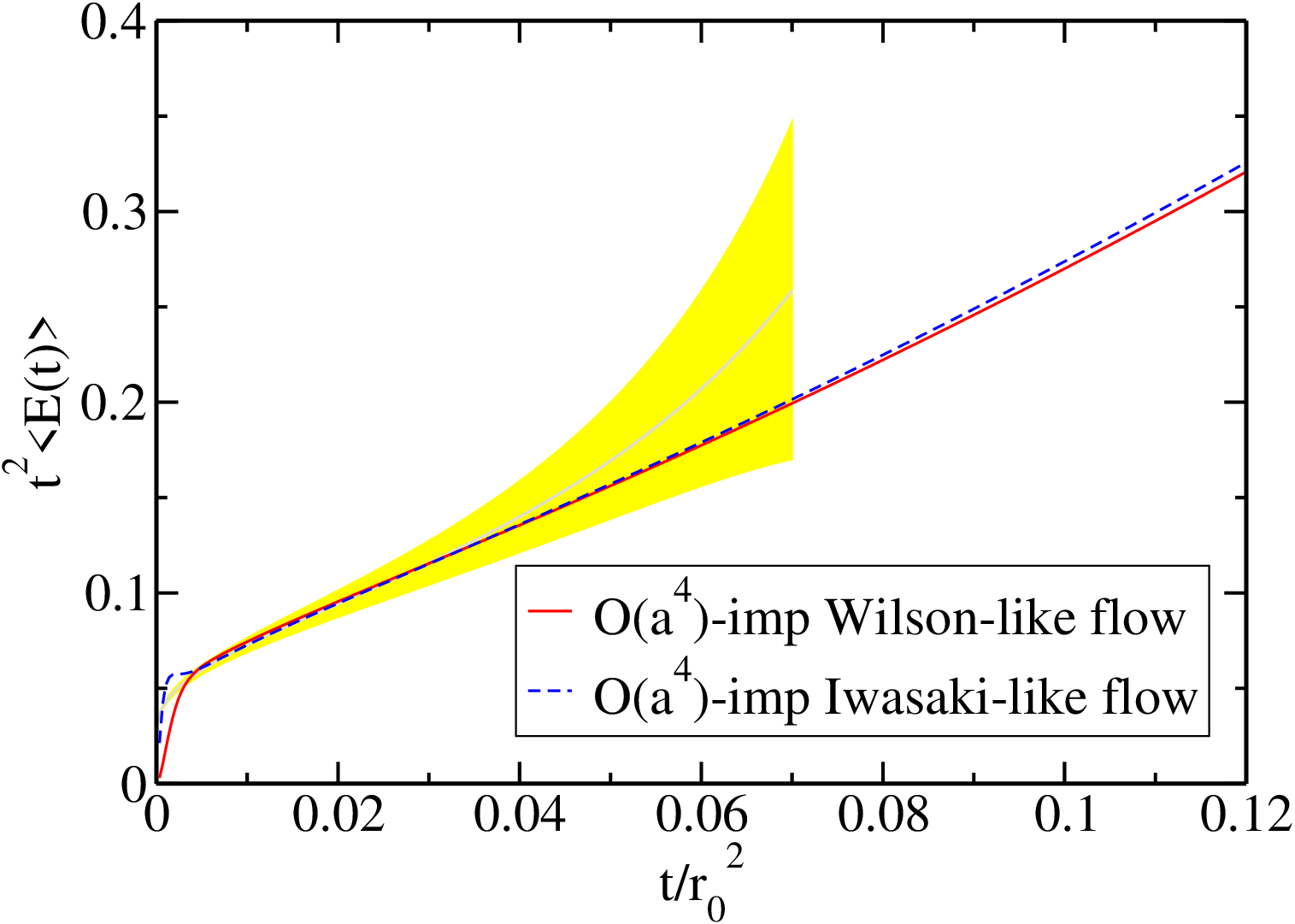}
\caption{The behavior of $t^2 \langle E (t) \rangle$ obtained 
 from tree-level $\mathcal{O}(a^4)$ improved flows as functions of $t/r_0^2$.
Three panels show results calculated at $\beta=5.96$ (left), $\beta=6.17$ (center) and
$\beta=6.42$ (right).}
 \label{fig:o4imp}
 \end{figure}

Figure~\ref{fig:o4imp} displays how the observed difference between the Wilson-type flow and the 
Iwasaki-type flow in the large $t$ regime can change when the lattice spacing decreases.
From the left panel to the right panel, the corresponding values of the lattice spacing in our simulations at given $\beta$ are going from a coarser to a finer lattice spacing. Clearly, the difference becomes diminished as the lattice spacing decreases.
Therefore, the difference stems from some remaining discretization errors. 
From these observations, we deduces that non-negligible $\mathcal{O}(g^2 a^2)$ corrections beyond the tree-level discretization effects sets in the large $t$ regime, where the renormalized coupling becomes large.
We then remark that the reference scale $t_0$ is determined at around $t/r_0^2\approx 0.1$ may suffer from rather large $\mathcal{O}(g^2 a^2)$ errors when the lattice spacing is coarse as large as $a\approx 0.1$ fm.

\section{Summary}
We have studied several types of tree-level improvement on the Yang-Mills gradient flow
in order to reduce the lattice discretization errors on $\langle E (t) \rangle$ 
in line with Ref.~\cite{Fodor:2014cpa}. 
For this purpose, the rectangle term is included in both the flow and gauge action in the minimal way.
We propose two-types of tree-level $\mathcal{O}(a^4)$ improved flow using the linear combination of  
two types of $\langle E (t) \rangle$ given by the plaquette- and clover-type definitions
and also perform numerical simulations for testing our proposal in the case of the Wilson plaquette gauge action for the configuration generation. 

Our numerical results have showed that 
tree-level lattice discretization errors on the quantity of $t^2\langle E (t) \rangle$ 
are certainly controlled in the small $t$ regime by both tree-level $\mathcal{O}(a^4)$ improved Wilson-like and Iwasaki-like flows. On the other hand, the values of $t^2\langle E (t) \rangle$ in the large $t$ regime
are different between the results given by two optimal flows leading to the same $\mathcal{O}(a^4)$ 
improvement at tree-level. However, the difference becomes diminished as the lattice spacing decreases.
We conclude that the remaining discretization errors found in the
large $t$ regime are supposed to stem from $\mathcal{O}(g^2 a^2)$ corrections 
beyond the tree-level discretization effects, since for the larger value of $t$ the higher order corrections at tree level become negligible due to powers of $a^2/t$ and meantime the renormalized coupling $g^2$ at a scale 
of $1/\sqrt{8t}$ becomes large. 

 We would like to thank the members of the FlowQCD Collaboration 
 (T. Hatsuda, T. Iritani, E. Itou, M. Kitazawa and H. Suzuki) for helpful suggestions and fruitful discussions.
 This work is in part based on Bridge++ code (http://bridge.kek.jp/Lattice-code/) and numerical calculations were  
 partially carried out on supercomputer SR16000 at YITP, Kyoto University.

\end{document}